\documentclass[10pt,a4papper]{article}
\usepackage[utf8]{inputenc}
\usepackage{indentfirst}
\usepackage[]{graphicx}
\usepackage{geometry} 
\begin{document}
	\title{\textbf{On the mixing angle of the vector mesons $\omega(782)$ and $\phi(1020)$}}
	\author{M. K. Volkov$^{1}$\footnote{volkov@theor.jinr.ru}, A. A. Pivovarov$^{1}$\footnote{tex$\_$k@mail.ru}, K. Nurlan$^{1,2,3}$\footnote{nurlan.qanat@mail.ru}\\
		\small
		\emph{$^{1}$BLTP, Joint Institute for Nuclear Research, Dubna, 141980, Russia}\\
		\small
		\emph{$^{2}$Institute of Nuclear Physics, Almaty, 050032, Kazakhstan}\\
		\small
		\emph{$^{3}$Eurasian National University, Nur-Sultan, 01008, Kazakhstan}}
	\date{}
	\maketitle
	\small
	
\begin{abstract}
	In the present work, the mixing angle of the vector $\omega(782)$ and $\phi(1020)$ mesons is estimated in the framework of Nambu--Jona-Lasinio model. The decay $\phi \to \pi^{0} \gamma$ is considered as a basic process to determine this angle. The obtained value is compared with the results of the other authors. Besides, the width of the decay $\phi \to 3\pi$ and the cross-section of the process $e^{+}e^{-} \to \pi^{0} \phi$ are calculated by using this angle.
	
\end{abstract}

\large
\section{Introduction}
	The mixing of the isoscalar vector mesons $\omega(782)$ and $\phi(1020)$ plays an important role in the description of the different processes of meson interactions. In the pseudoscalar case, the singlet-octet mixing leading to the physical mesons $\eta$ and $\eta^{'}(958)$ has been described by many authors. For example, in the works \cite{Klimt:1989pm,Vogl:1991qt,Volkov:1998ax}, in the framework of Nambu--Jona-Lasinio (NJL) model \cite{Vogl:1991qt,Nambu:1961tp,Ebert:1982pk,Volkov:1984kq,Volkov:1986zb,Ebert:1985kz,Klevansky:1992qe,Volkov:1993jw,Ebert:1994mf}, it is shown that the 't Hooft interaction allows one to describe the masses of pseudoscalar mesons and their singlet-octet mixing angle. In the vector case, as a mechanism of mesons $\omega(782)$ and $\phi(1020)$ mixing one may consider their interaction via the kaon loops. This mechanism in the framework of the hidden local symmetry Lagrangian was described in \cite{Benayoun:2007cu,Benayoun:2009im}. In the first work, the value $4.6^{\circ}$ was obtained at the energies of the $\phi(1020)$ meson mass. In the second one, this value was changed to $3.84^{\circ}$. The effect of the $\omega(782)$ and $\phi(1020)$ mesons mixing has been considered in the numerous other theoretical works. In the work \cite{Klingl:1996by} this angle was defined by using the process $\phi \to \pi^{0} \gamma$ calculated with the chiral $SU(3)$ symmetric Lagrangian. The value $3.3^{\circ}$ was obtained. In the work \cite{Kucukarslan:2006wk}, the authors received the angle $3.4 \pm 0.3^{\circ}$ in the framework of the Chiral perturbation theory. In the work \cite{Choi:2015ywa}, in the light-front quark model with the QCD-motivated effective Hamiltonian including the hyperfine interaction, the value $5.2^{\circ}$ was obtained. The collaboration KLOE experimentally received  the value $3.32 \pm 0.09^{\circ}$ \cite{Ambrosino:2009sc}.
	
	In the present work, we do not consider the nature of these mesons mixing in detail. However we make the estimation of this angle by using different decays calculated in the NJL model. As a basic process, like in the work \cite{Klingl:1996by}, we use the decay $\phi \to \pi^{0} \gamma$, because it has been measured with enough precision. First of all we calculate the mesons $\omega(782)$ and $\phi(1020)$ mixing angle by using this decay in the framework of NJL model. Than we apply this angle to investigate the processes $\phi \to 3\pi$ and $e^{+}e^{-} \to \pi^{0} \phi$.
	
\section{The interaction Lagrangian of the extended NJL model}
	In the NJL model, a fragment of u and d quark part of the quark-meson interaction Lagrangian, containing the vertices we need, takes the form \cite{Volkov:1986zb}:
	\begin{eqnarray}
		\Delta L_{int} & = &
		\bar{q} \left[ \frac{g_{a_{1}}}{2} \gamma^{\mu} \gamma^{5} \sum_{j=\pm,0} \lambda_{j}^{a_{1}} a^{j}_{1\mu} + \frac{g_{\phi}}{2} \sin(\alpha) \gamma^{\mu} \lambda^{\phi} \phi_{\mu} + \frac{g_{\omega}}{2} \cos(\alpha) \gamma^{\mu} \lambda^{\omega} \omega_{\mu} + i g_{\pi} \gamma^{5} \sum_{j=\pm,0} \lambda_{j}^{\pi} \pi^{j}\right]q,
	\end{eqnarray}
	where $q$ and $\bar{q}$ are the u and d quark fields with the constituent masses $m_{u} \approx m_{d} = m = 280$~MeV, s quarks do not appear here since they do not participate in the studied processes. The sine in the second term and the cosine in the third term take into account the mixing of the $\omega(782)$ and $\phi(1020)$ mesons.
	
	The coupling constants are:
	\begin{eqnarray}
	\label{Couplings}
		g_{a_{1}} = g_{\phi} = g_{\omega} = \left(\frac{2}{3}I_{2}\right)^{-1/2} \approx 6.14, &\quad& g_{\pi} = \left(\frac{4}{Z_{\pi}}I_{2}\right)^{-1/2} \approx 3.01,
	\end{eqnarray}
	where
	\begin{eqnarray}
		Z_{\pi} & = & \left(1 - 6\frac{m^{2}}{M^{2}_{a_{1}}}\right)^{-1} \approx 1.45.
	\end{eqnarray}
	
	Here $Z_{\pi}$ is the additional renormalization constant appearing in the $\pi - a_{1}$ transition, $M_{a_{1}} = 1230 \pm 40$~MeV is the mass of the axial-vector $a_{1}(1260)$ meson \cite{Tanabashi:2018oca}.
	
	The integral appearing in the quark loops as a result of renormalization of the Lagrangian has the form
	\begin{eqnarray}
		I_{2} = -i\frac{N_{c}}{(2\pi)^{4}}\int\frac{\Theta(\Lambda_{4}^{2} + k^2)}{(m^{2} - k^2)^{2}}\mathrm{d}^{4}k,
	\end{eqnarray}
	where $\Lambda_{4} = 1.26$~GeV  is the four-dimensional cutoff parameter \cite{Volkov:1986zb}, $N_{c} = 3$ is the number of colors in QCD.
	The matrices $\lambda$ are linear combinations of Pauli matrices: 
	
	\begin{eqnarray}
		\lambda_{0}^{a_{1}} = \lambda_{0}^{\pi} = \left(\begin{array}{cc}
		1 & 0 \\
		0 & -1 \\
		\end{array}\right), && 
		\lambda_{-}^{a_{1}} = \lambda_{-}^{\pi} = \sqrt{2} \left(\begin{array}{ccc}
		0 & 0 \\
		1 & 0 \\
		\end{array}\right), \nonumber\\
		\lambda^{\phi} = \lambda^{\omega} = \left(\begin{array}{cc}
		1 & 0 \\
		0 & 1 \\
		\end{array}\right), &&
		\lambda_{+}^{a_{1}} = \lambda_{+}^{\pi} = \sqrt{2} \left(\begin{array}{ccc}
		0 & 1 \\
		0 & 0 \\
		\end{array}\right).
	\end{eqnarray}
	
\section{The process $\phi \to \pi^{0} \gamma$}
\begin{figure}[h]
		\center{\includegraphics[scale = 0.7]{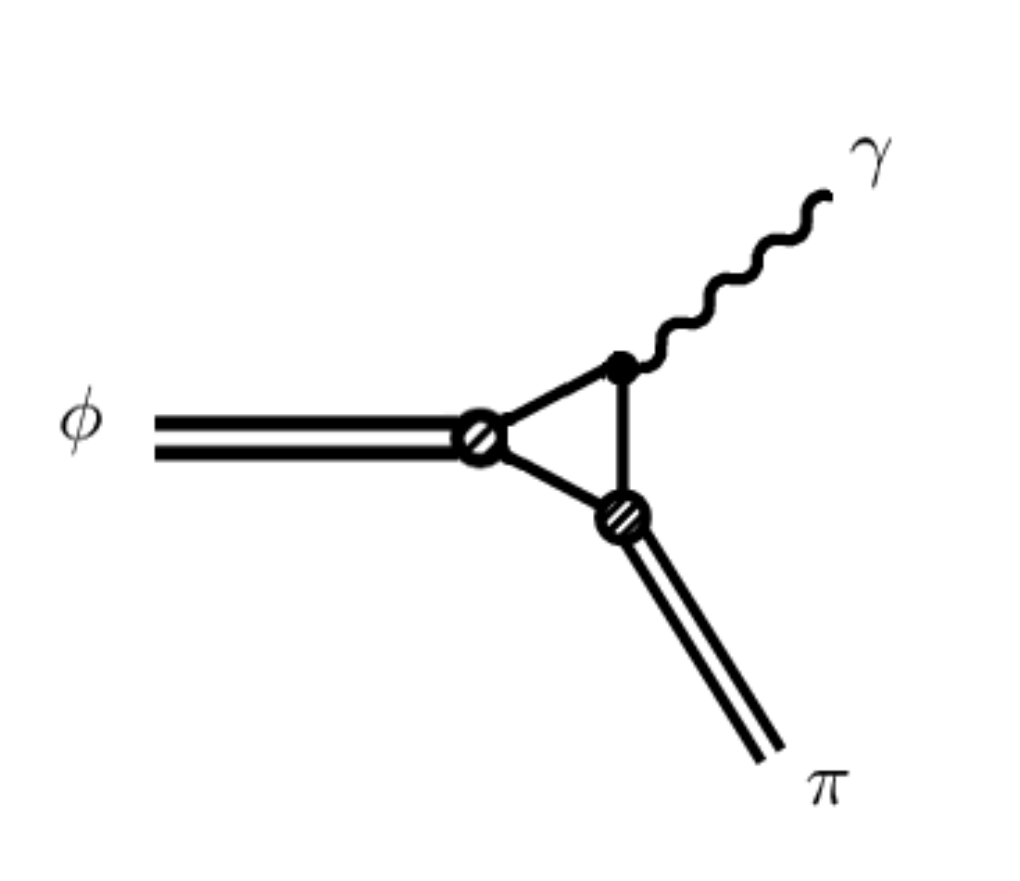}}
		\caption{The diagramm of the decay $\phi \to \pi^{0} \gamma$.}
		\label{phi_pigamma}
	\end{figure}

	The diagram of the process $\phi \to \pi^{0} \gamma$ is presented in the Fig.~\ref{phi_pigamma}.
	
	The amplitude of the considered process in the NJL model takes the form:
	\begin{eqnarray}
		\mathcal{M}(\phi \to \pi^{0} \gamma) & = & \frac{3}{4} \frac{\sqrt{\alpha_{em}}}{\pi^{3/2} F_{\pi}} g_{\phi} \sin(\alpha) e^{\mu \nu \lambda \delta} e_{\mu}(p_{\phi}) e_{\nu}^{*}(p_{\gamma}) p_{\pi \lambda} p_{\gamma \delta},
	\end{eqnarray}
	where $\alpha_{em}$ is the electromagnetic constant, $F_{\pi} = \frac{m}{g_{\pi}} \approx 93$ MeV is the pion decay constant, $e_{\mu}^{*}(p_{\phi})$ and $e_{\nu}^{*}(p_{\gamma})$ are the polarisation vectors of the meson $\phi(1020)$ and the photon. The similar amplitude for the process $\omega \to \pi^{0} \gamma$ was obtained in the NJL model in \cite{Volkov:1986zb}.
	
	The experimental value of this decay width \cite{Tanabashi:2018oca}:
	\begin{eqnarray}
		\Gamma(\phi \to \pi^{0} \gamma)_{exp} = 5.5 \pm 0.2 \textrm{ keV}.
	\end{eqnarray}
	
	By using this known experimental value one can fix the mixing angle of the mesons $\omega(782)$ and $\phi(1020)$. The result is $\alpha = 3.1^{\circ}$.
	
\section{The process $\phi \to 3\pi$}	
\begin{figure}[h]
		\center{\includegraphics[scale = 0.6]{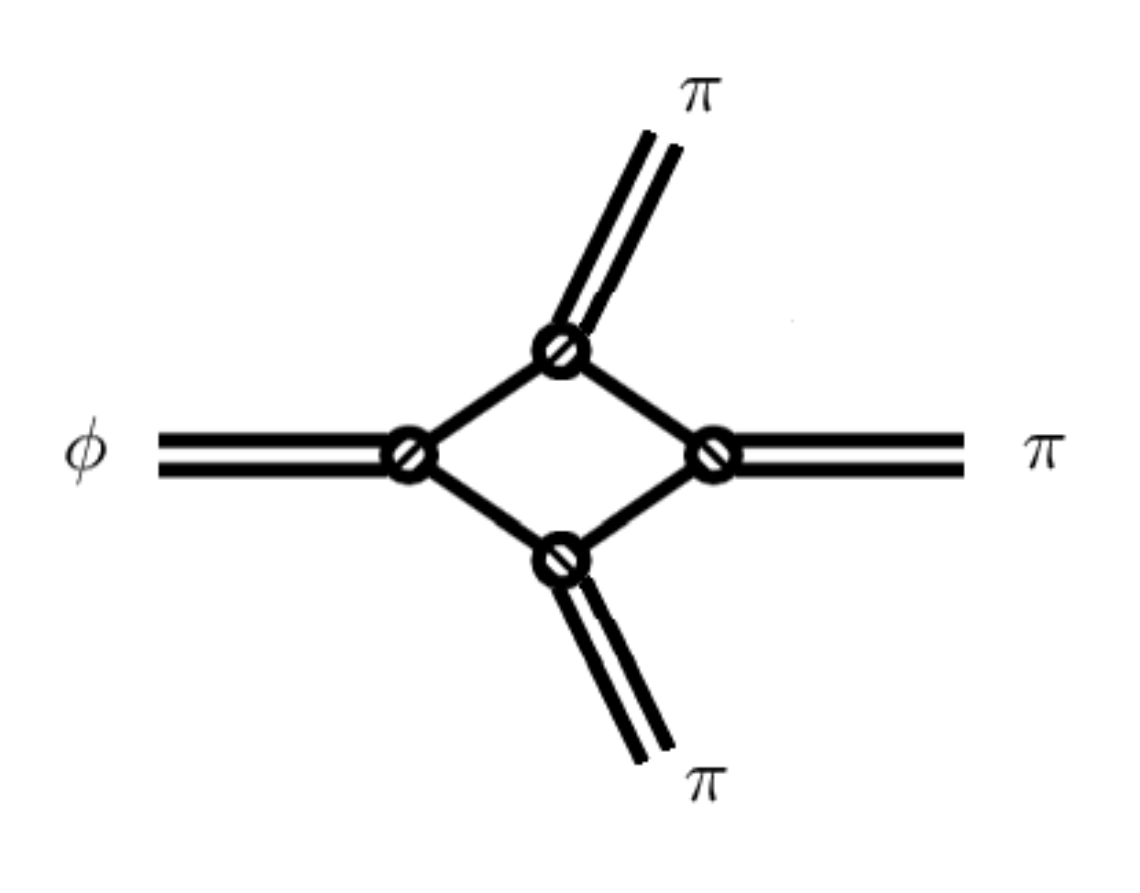}}
		\caption{The box diagram of the decay $\phi \to 3 \pi$.}
		\label{Box}
	\end{figure}

\begin{figure}[h]
		\center{\includegraphics[scale = 0.7]{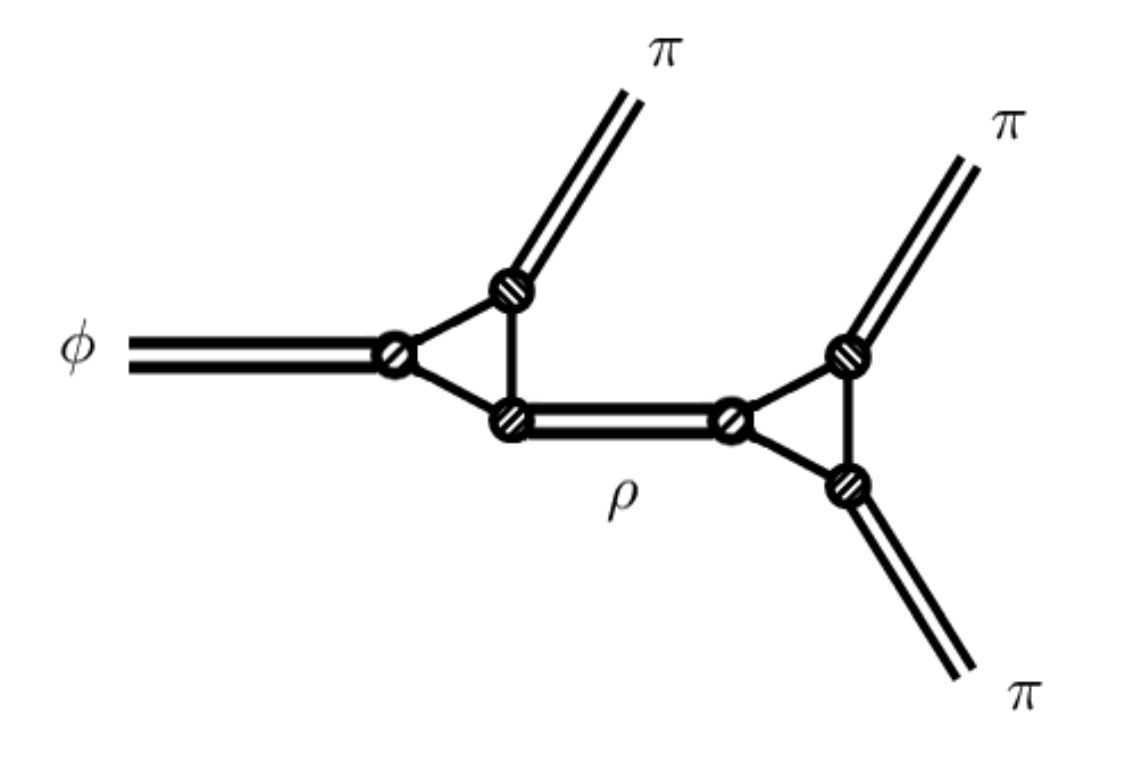}}
		\caption{The diagram of the decay $\phi \to 3 \pi$ with the intemediate $\rho$ meson.}
		\label{Triangle}
	\end{figure}

	In the NJL model, the decay $\phi \to 3 \pi$ is described with two types of the anomalous quark diagrams: anomalous box quark diagram (Fig. \ref{Box}) and diagram with the intermediate $\rho$ meson which connects the anomalous triangle diagram with the vertex $\rho \to \pi \pi$ (Fig. \ref{Triangle}).
	
	This process is similar to the well-known decay $\omega \to 3 \pi$ investigated in the NJL model recently in \cite{Osipov:2020hjd} and can be obtained from it due to $\omega - \phi$ mixing:
	\begin{eqnarray}
	\label{amplitude}
		\mathcal{M}(\phi \to 3\pi) = - \frac{3}{4} \frac{g_{\phi}\sin(\alpha)}{\pi^2{F_{\pi}}^3} \biggl[ b + {g_{\rho}}^2{F_{\pi}}^2 \sum_{i=+,-,0} \frac{1}{M^2_{\rho} - q_{i}^2-i \sqrt{q_{i}^2} \Gamma_{\rho}(q_{i}^2)}	  \biggl] e^{\mu \nu \lambda \delta} e_{\mu}(p_{\phi}) p_{0 \nu} p_{+ \lambda} p_{- \delta}.
	\end{eqnarray}
	where $p_{0}, p_{-},p_{+}$ are the pions momenta, $p_{\phi}$ is the momentum of the $\phi$ meson, $q_{i} = p_{\phi} - p_{i}$ ($i = +,-,0$) are the momenta of the intermediate $\rho$ mesons, $M_\rho = 775.49$ MeV \cite{Tanabashi:2018oca} is the $\rho$ meson mass. The first term of the amplitude describes the contribution of the box diagram. The $\pi-a_1$ transitions on the pion lines lead to the following form of the constant $b$:	
	\begin{eqnarray}
		b = 1-\frac{3}{a}+\frac{3}{2a^2}+\frac{1}{8a^3},
	\end{eqnarray}   
	where $a = 1.84$ \cite{Osipov:2020hjd}. The second, the third and the fourth terms describe one, two and three $\pi-a_1$ transitions respectively. The decay width of the intermediate $\rho$ meson depends on the momentum:	
	\begin{eqnarray}
		\Gamma_{\rho}(q^2) = \frac{{g_{\rho}}^2 ({q}^2 - 4{M_{\pi}}^2)^{3/2}}{48{\pi} {q}^2}.
	\end{eqnarray}  
	
	The result for the full width of the decay $\phi \to 3 \pi$ in the NJL model is
	\begin{eqnarray}
		\Gamma (\phi \to  3\pi) = 0.67 \textrm{ MeV},
	\end{eqnarray} 
	
	The experimental value \cite{Akhmetshin:1995vz}:	
	\begin{eqnarray}
		\Gamma (\phi \to  3\pi)_{exp} = (0.684 \pm 0.036) \textrm{ MeV},
	\end{eqnarray}

\section{The process $e^+e^- \to \phi \pi^{0}$}
\begin{figure}[h]
		\center{\includegraphics[scale = 0.6]{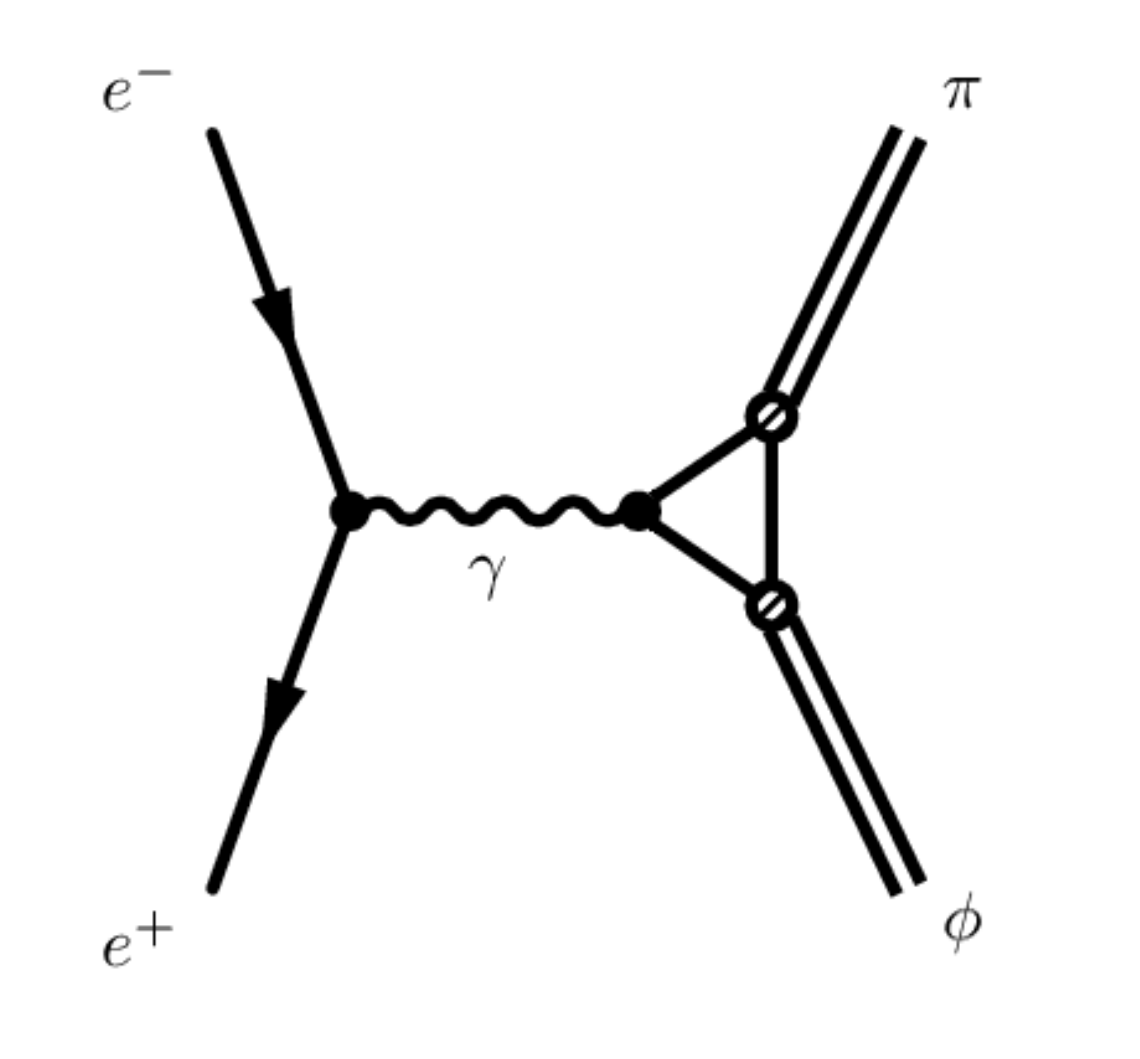}}
		\caption{The contact diagram of the process $e^+e^- \to \phi \pi^{0}$.}
		\label{Contact}
	\end{figure}

\begin{figure}[h]
		\center{\includegraphics[scale = 0.7]{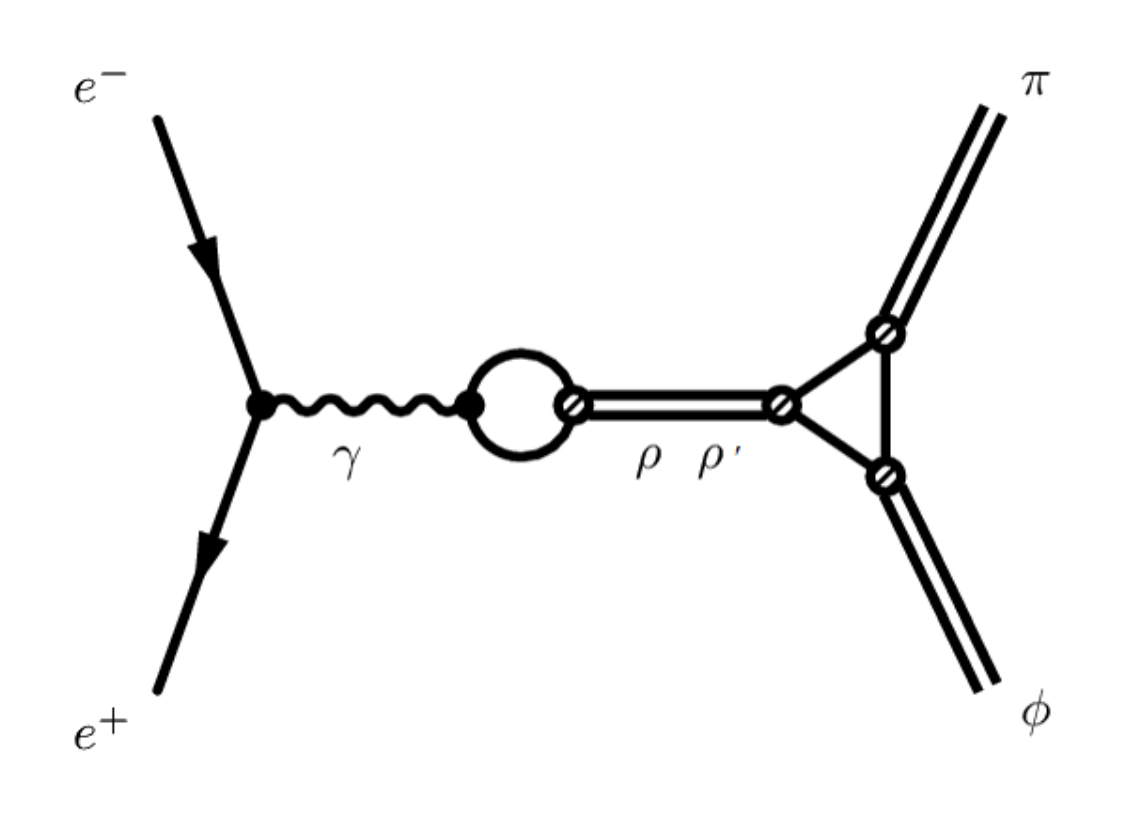}}
		\caption{The diagram of the process $e^+e^- \to \phi \pi^{0}$ with the intermediate $\rho, \rho'$ mesons.}
		\label{Intermediate}
	\end{figure}

	The diagrams of the processes $e^+e^- \to \phi \pi^{0}$  are shown in Figs.~\ref{Contact},~\ref{Intermediate}.	
	
	
	The amplitude of the process $e^+e^- \to \phi \pi^{0}$ is calculated in the extended NJL model \cite{Volkov:1996br,Volkov:1996fk, Volkov:2017arr} (see Appendix), because the meson $\rho(1450)$ considered as the first radially excited state can contribute to this process. As a result, the amplitude takes the form:
	
	\begin{equation}
		\mathcal{M}(e^+e^- \to \phi \pi^{0}) = \frac{8\pi\alpha_{em}}{s}m \sin(\alpha) (T_{W} + T_{\rho} + T_{\rho'}) l_{\mu} \epsilon^{\mu \nu \lambda \delta} e_{\nu}^{*}(p_{\phi}) p_{\phi \lambda}  p_{\pi \delta},
	\end{equation}
	where $s=(p_{e^+}+p_{e^-})^2$, $l^{\mu}=\bar e \gamma^\mu e$ is the lepton current. 
	
	The terms corresponding to contributions from the contact diagram and the diagram with the intermediate $\rho$ meson are
	
	\begin{equation}
		T_{W} = g_{\pi} I^{\phi}_{3},
	\end{equation}
	
	\begin{equation}
		T_{\rho} = \frac{C_{\rho} g_{\pi} I^{\rho\phi}_{3}}{g_{\rho}}\frac{s}{M_\rho^2 - s - i\sqrt{s}\Gamma_\rho(s)}.
	\end{equation}
	
	The contribution of the amplitude with the intermediate meson $\rho(1450)$:
	
	\begin{equation}
		T_{\rho'} = \frac{C_{\rho'} g_{\pi} I^{\rho'\phi}_{3}}{g_{\rho}}\frac{s}{M_{\rho'}^2 - s - i\sqrt{s}\Gamma_{\rho'}(s)}.
	\end{equation}
	
	The constants $C_{\rho}$, $C_{\rho'}$ appear in the quark loops of photon transition into the intermediate meson:
	
	\begin{eqnarray}
		C_{\rho} & = & \frac{1}{\sin\left(2\theta_{\rho}^{0}\right)}\left[\sin\left(\theta_{\rho} + \theta_{\rho}^{0}\right) +	R_{\rho}\sin\left(\theta_{\rho} - \theta_{\rho}^{0}\right)\right], \nonumber \\
		C_{\rho}^{'} & = & \frac{-1}{\sin\left(2\theta_{\rho}^{0}\right)}\left[\cos\left(\theta_{\rho} + \theta_{\rho}^{0}\right) + R_{\rho}\cos\left(\theta_{\rho} - \theta_{\rho}^{0}\right)\right].
	\end{eqnarray}
	
	The integrals, which appear in the quark loops are
	
	\begin{eqnarray}
	\label{DiffIntegral}
		I_{3}^{\phi} & = &
		-i\frac{N_{c}}{(2\pi)^{4}}\int\frac{A_{\phi}({\bf k})}{(m^{2} - k^2)^{3}} \Theta(\Lambda_{3}^{2} - {\bf k}^2)	\mathrm{d}^{4}k, \nonumber\\
		I_{3}^{\rho\phi} & = &
		-i\frac{N_{c}}{(2\pi)^{4}}\int\frac{A_{\rho}({\bf k}) A_{\phi}({\bf k})}{(m^{2} - k^2)^{3}} \Theta(\Lambda_{3}^{2} - {\bf k}^2)	\mathrm{d}^{4}k, \nonumber\\
		I_{3}^{\rho^{'}\phi} & = &
		-i\frac{N_{c}}{(2\pi)^{4}}\int\frac{B_{\rho}({\bf k}) A_{\phi}({\bf k})}{(m^{2} - k^2)^{3}} \Theta(\Lambda_{3}^{2} - {\bf k}^2)	\mathrm{d}^{4}k,
	\end{eqnarray}
	where $\Lambda_{3}$ is the three-dimensional cutoff parameter. The parameters used here are defined in the Appendix.
	
\begin{figure}[h!]
		\caption{ The cross-section of the process $e^+e^- \to \phi \pi^{0}$. The experimental points are taken from the work of the BaBar collaboration \cite{Aubert:2007ym}.}
		\center{\includegraphics[scale = 0.5]{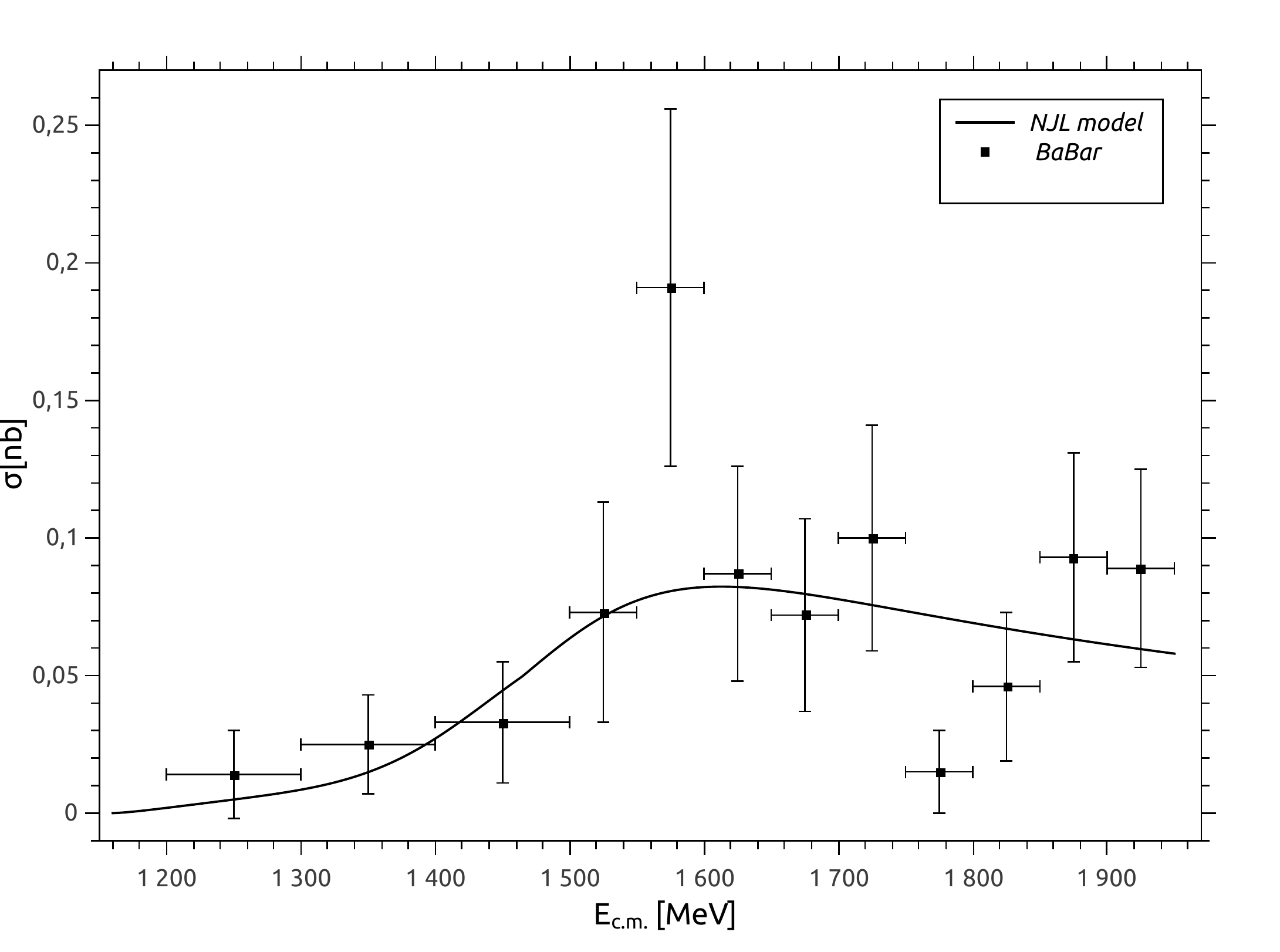}}
		\label{CrossSection}
	\end{figure}

	A comparison of the cross-section of the process $e^+e^- \to \phi \pi^{0}$ with the experimental data is presented in Fig. \ref{CrossSection}. The obtained results are in satisfactory agreement with the experimental data.
	
\section{Conclusion}
	The consideration of the mesons $\omega(782)$ and $\phi(1020)$ mixing angle is necessary to describe different processes. This angle is determined by various authors in different ways. In the present work, by using the decays $\phi \to \pi^{0} \gamma$, $\phi \to 3\pi$ and $e^{+}e^{-} \to \pi^{0} \phi$ calculated in the NJL model the result $3.1^\circ$ have been obtained. The error of this model is determined by the chiral symmetry breaking effects and by the model parameters uncertainty. The error caused by the first reason is $\frac{M_{\pi}}{M_{p}} \approx 2 \%$. The total accuracy of the NJL model may be estimated by comparing the previous results obtained in the framework of this model for different processes with the experimental data. This accuracy is about 10\%. Thus, our estimation for the given mixing angle is $3.1 \pm 0.3^\circ$. The angle $3^{\circ}$ has been used in our earlier works, particularly, in the process $e^{+}e^{-} \to \pi^{0} \gamma$ \cite{Arbuzov:2011fv}, and has led to a good agreement with the experiment. This result is close to the results obtained in \cite{Klingl:1996by, Kucukarslan:2006wk}, where the chiral symmetry has been used. Further theoretical and experimental researches aimed at an investigation of the nature of this mixing are of interest.
    
\section*{Appendix. The Lagrangian of the extended NJL model}    
	In the extended NJL model, the part of the quark–meson interaction Lagrangian referring to the mesons involved in the process under consideration has the form \cite{Volkov:1996br,Volkov:1996fk,Volkov:2017arr}:
	\begin{eqnarray}
	\Delta L_{int} & = &
		\bar{q} \left[ \frac{1}{2} \gamma^{\mu}\gamma^{5} \sum_{j=\pm,0} \lambda_{j}^{a_{1}} \left(A_{a_{1}}a^{j}_{1\mu} + B_{a_{1}}a^{'j}_{1\mu}\right)
		+ \frac{1}{2} \sin\alpha \gamma^{\mu} \lambda^{\phi} \left(A_{\phi}\phi_{\mu} + B_{\phi}\phi^{'}_{\mu}\right) \right. \nonumber\\
		&& + \left. \frac{1}{2} \gamma^{\mu} \sum_{j=\pm,0} \lambda_{j}^{\rho} \left(A_{\rho}\rho^{j}_{\mu} + B_{\rho}\rho^{'j}_{\mu}\right) + i \gamma^{5} \sum_{j = \pm,0} \lambda_{j}^{\pi} \left(A_{\pi}\pi^{j} + B_{\pi}\pi^{'j}\right)\right]q,
	\end{eqnarray}
	where the excited meson states are marked with prime,
	\begin{eqnarray}
	\label{verteces}
		A_{M} & = & \frac{1}{\sin(2\theta_{M}^{0})}\left[g_{M}\sin(\theta_{M} + \theta_{M}^{0}) +
		g_{M}^{'}f_{M}(k_{\perp}^{2})\sin(\theta_{M} - \theta_{M}^{0})\right], \nonumber\\
		B_{M} & = & \frac{-1}{\sin(2\theta_{M}^{0})}\left[g_{M}\cos(\theta_{M} + \theta_{M}^{0}) +
		g_{M}^{'}f_{M}(k_{\perp}^{2})\cos(\theta_{M} - \theta_{M}^{0})\right].
	\end{eqnarray}
	The subscript $M$ specifies the corresponding meson.
	
	The first radially excited states are introduced using the form factor $f\left(k_{\perp}^{2}\right) = \left(1 + d k_{\perp}^{2}\right)$. The slope parameter $d$ was obtained from the requirement of invariability of the quark condensate after including the radially excited meson states and it depends only on the quark content of the meson:
	
	\begin{eqnarray}
		&d_{uu} = -1.784 \times 10^{-6} \textrm{MeV}^{-2}.&
	\end{eqnarray}
	
	The transverse relative momentum of the inner quark-antiquark system can be represented as
	
	\begin{eqnarray}
		k_{\perp} = k - \frac{(kp) p}{p^2},
	\end{eqnarray}
	where $p$ is the meson momentum. In the rest system of a meson
	\begin{eqnarray}
		k_{\perp} = (0, {\bf k}).
	\end{eqnarray}
	Therefore, this momentum may be used as a three-dimensional one.
	
	The parameters $\theta_{M}$ are the mixing angles determined after diagonalization of the free Lagrangian for the ground and first radially excited states \cite{Volkov:1996fk,Volkov:2017arr}:	
	\begin{eqnarray}
		\theta_{a_{1}} = \theta_{\rho} = \theta_{\phi} = 81.8^{\circ}, &\quad& \theta_{\pi} = 59.48^{\circ}.
	\end{eqnarray}
	In addition, $\theta_{M}^{0}$ are auxiliary parameters introduced for convenience as
	\begin{eqnarray}
	\label{tetta0}
		&\sin\left(\theta_{M}^{0}\right) = \sqrt{\frac{1 + R_{M}}{2}},& \nonumber\\
		&R_{a_{1}} = R_{\rho} = R_{\phi} = \frac{I_{2}^{f_{uu}}}{\sqrt{I_{2}I_{2}^{f^{2}}}}, & \nonumber\\ & 
		R_{\pi} = \frac{I_{2}^{f}}{\sqrt{Z_{\pi}I_{2}I_{2}^{f^{2}}}},&
	\end{eqnarray}
	The integrals appearing in the quark loops as a result of renormalization of the Lagrangian are
	\begin{eqnarray}
		I_{2}^{f^{m}} =
		-i\frac{N_{c}}{(2\pi)^{4}}\int\frac{f^{m}({\bf k}^{2})}{(m^{2} - k^2)^{2}}\Theta(\Lambda_{3}^{2} - {\bf k}^2)
		\mathrm{d}^{4}k,
	\end{eqnarray}
	where $\Lambda_{3} = 1.03$~GeV is the three-dimensional cutoff parameter.
	
	Then
	\begin{eqnarray}
		&\theta_{a_{1}}^{0} = \theta_{\rho}^{0} = \theta_{\phi}^{0} = 81.5^{\circ}, \quad \theta_{\pi}^{0} = 59.12^{\circ}.& 
	\end{eqnarray}
	The constants $g_{a_{1}}$, $g_{\phi}$ and $g_{\pi}$ were defined in (\ref{Couplings}). The constant $g_{\rho}$ is the same as $g_{\phi}$. The constants appearing due to introducing of excited states have the following form:
	\begin{eqnarray}
		g_{a_{1}}^{'} = g_{\rho}^{'} = g_{\phi}^{'} = \left(\frac{2}{3}I_{2}^{f^{2}}\right)^{-1/2}, &\quad& g_{\pi}^{'} =  \left(4 I_{2}^{f^{2}}\right)^{-1/2}.
	\end{eqnarray}	
	
\section*{Acknowlegments}
	The authors are grateful to A. A. Osipov and A. B. Arbuzov for useful discussions.

\end{document}